\documentstyle[12pt,preprint,aps]{revtex}

\begin{document}

\title{The Saffman-Taylor problem on a sphere}

\author{Fernando Parisio, Fernando Moraes, Jos\'e A. Miranda}
\address{Laborat\'{o}rio de F\'{\i}sica Te\'{o}rica e Computacional,
Departamento de F\'{\i}sica,\\ Universidade Federal de Pernambuco,
Recife, PE  50670-901 Brazil}

\author{Michael Widom}
\address{Department of Physics, Carnegie Mellon University, 
Pittsburgh, PA  15213 USA}
\date{\today}
\maketitle

\begin{abstract}
The Saffman-Taylor problem addresses the morphological instability of
an interface separating two immiscible, viscous fluids when they move
in a narrow gap between two {\it flat} parallel plates (Hele-Shaw
cell).  In this work, we extend the classic Saffman-Taylor situation,
by considering the flow between two {\it curved}, closely spaced,
concentric spheres (spherical Hele-Shaw cell). We derive the
mode-coupling differential equation for the interface perturbation
amplitudes and study both linear and nonlinear flow regimes.  The
effect of the spherical cell (positive) spatial curvature on the shape
of the interfacial patterns is investigated. We show that stability
properties of the fluid-fluid interface are sensitive to the 
curvature of the surface. In particular, it is found
that positive spatial curvature inhibits finger
tip-splitting. Hele-Shaw flow on weakly negative, curved surfaces is
briefly discussed.
\end{abstract}
\pacs{PACS number(s):47.20.-k, 68.10.-m, 47.54.+r, 02.40.-k}

\section{Introduction}
\label{intro}

Formation and evolution of dynamic structures is an exciting area of
nonlinear phenomenology. Of particular practical and theoretical
interest is the hydrodynamic pattern formation of the growing
interface between two fluids. One of the best studied
pattern-formation systems of this type is the Saffman-Taylor
problem~\cite{Saf,Rev}: it addresses motion of two viscous, immiscible
fluids in the narrow space between two parallel, flat plates known as
a Hele-Shaw cell. When a fluid of low viscosity displaces a fluid of
higher viscosity, the interface between them becomes unstable and
deforms. Dynamic competition leads to the formation of fingering
structures.

Experiments and theory focus on two principal geometries: (i)
rectangular~\cite{Saf,Rev} and (ii) radial~\cite{Bat,Wil,Pat}. In
rectangular cells the unperturbed interface is straight and the
unperturbed flow is uniform and parallel to cell walls. In the radial
case the unperturbed interface is circular with the less viscous fluid
pumped into the more viscous one at a point and the flow radially
outward.  For both situations, the initial development of the
interface instability tracks the predictions of linear stability
theory~\cite{Saf,Rev,Bat,Wil,Pat}.  After the initial surface
deformation, as the unstable modes of perturbation grow, they become
coupled in a weakly nonlinear stage of
evolution~\cite{Mir,MW}. Finally, the system evolves to a complicated
late stage, characterized by formation of fingering structures, in
which nonlinear effects dominate~\cite{Rev}. As a result, beautiful
finger-bubble undulated structures are formed in rectangular cells,
while visually striking, fan-like, branched patterns rise in radial
flow. Spreading, splitting and competition are the three basic growth
mechanisms of the viscous fingering process~\cite{Rev}. In this work
we are particularly interested in tip-splitting events.

Despite the extensive experimental and theoretical work on the
Saffman-Taylor problem in both radial and rectangular setups, the
majority of the studies focus on flow in {\it flat} Hele-Shaw
cells. Curiously, the dynamic behavior for flow in flat cells is
described by the very same set of equations as those for flow in
porous media~\cite{Rev}, which is indeed a highly non-flat
environment, characterized by voids and curved internal surfaces. Even
though viscous fingering is not restricted to occur between flat
surfaces, the study of the Saffman-Taylor problem on curved surfaces
has been practically neglected. One exception is the experimental work
by Zhao and Maher~\cite{Zhao}, which considers flow in a cylindrical
Hele-Shaw cell with a large radius of curvature. The flow is performed 
parallel to the cell axis. They used the cylindrical cell as an experimental
realization of a flat rectangular Hele-Shaw cell with periodic
boundary conditions.  Their experiments showed that instabilities in
cells with periodic boundary conditions are qualitatively similar to
instabilities in cells with physical sidewalls. The authors did not 
explore the influence of cell curvature.

On the theoretical side, Entov and Etingof~\cite{Entov} considered the
general mathematical problem of viscous flow in non-planar Hele-Shaw
cells, in the zero-surface-tension limit.  They applied conformal
mapping techniques to derive a class of explicit solutions for the
shape of the fluid domain under study. As in reference~\cite{Zhao},
the authors in~\cite{Entov} were not interested in the influence of
the cell curvature in the shape of the interfacial patterns.
 
Spatial curvature has been a relatively overlooked feature in the
study of other pattern formation systems as well. Only recently
researchers started to investigate how the properties of the patterns
could be affected by the curvature of the surface in which such
structures evolve~\cite{Lev1,Lev2,Hyde,Zyk,Var,Desai}.  In the early
1990's Levine and collaborators studied coarsening of two-dimensional
foams~\cite{Lev1} and grain growth~\cite{Lev2} on curved
surfaces. They showed that the stability properties of such froth
bubbles and grains depended on the curvature of the surface.
Ref.~\cite{Hyde} reviews the dependence of many chemical and
biochemical surface processes on curvature.  Numerical studies of
reaction-diffusion systems in curved spaces examine the evolution of
spiral waves~\cite{Zyk} and the occurence of Turing patterns on a
sphere~\cite{Var}. These numerical simulations~\cite{Zyk,Var} indicate
that curvature imposes geometrical restrictions on the shape of the
patterns. Finally, Schoenborn and Desai~\cite{Desai} studied the
intra-surface kinetics of phase ordering on curved surfaces.

For the Saffman-Taylor problem, the interplay between Hele-Shaw cell
curvature and interfacial pattern formation is largely unexplored in
the present literature. However, in a recent mode-coupling analysis of
radial flow in flat cells, Miranda and Widom~\cite{MW} suggested that
cell curvature could be used as a control parameter to regulate the
tendency towards finger tip-splitting. A thorough investigation of the
relationship between cell curvature and the fluid-fluid interface
dynamics still needs to be addressed. In this work we begin such
investigations, focusing on the Saffman-Taylor problem on a sphere. 
Gravity effects have been studied in reference~\cite{grav}.

The study of viscous flow in a nonplanar Hele-Shaw cell is of interest
for both scientific and practical reasons.  On the scientific level,
the influence of spatial curvature on hydrodynamic flow is a matter of
fundamental interest.  It also provides a simple mathematical model to
describe more general situations involving the filling of a thin
cavity between two walls of a given shape with fluid. On the practical
level, it may have applications in a number of industrial,
manufacturing processes, ranging through pressure moulding of molten
metals and polymer materials~\cite{Rich}, and formation of coating
defects in drying paint thin films~\cite{Schwartz}.

The outline of the work is the following: section~\ref{derivation}
defines Hele-Shaw flow between concentric, thin spherical shells,
considering fluid injection (withdrawal) at the north (south) pole. We
derive a differential equation describing the early nonlinear
evolution of the interface modes. In section~\ref{discussion}, we
interpret results obtained in section~\ref{derivation} and investigate
both the linear and weakly nonlinear evolution of the system. The flat
space limit of infinite radius of curvature is examined in
section~\ref{flat}. Section~\ref{linear} discusses the linear growth
rates of unstable modes and relates these to the degree of lattitude
of the unperturbed interface.  Nonlinear analysis in
section~\ref{nonlinear} concentrates on the effect of cell curvature
on finger tip-splitting.  We show that positive spatial curvature
inhibits splitting.  Flow on the northern and southern hemispheres are
contrasted, and a symmetry-breaking behavior is detected: for the
southern hemisphere tip-splitting is replaced by finger
tip-sharpening. Flow on weakly negative curved surfaces is briefly
discussed. Section~\ref{conclude} presents our final remarks. An
appendix derives Darcy's law for flow between concentric spheres.

\section{The mode coupling differential equation} 
\label{derivation}

Consider two immiscible, incompressible, viscous fluids, flowing in a
narrow gap of thickness $b$, between two concentric, thin spherical
shells (see figure 1). We name this device the {\it spherical}
Hele-Shaw cell. Assume that $b$ is smaller than any other length scale
in the problem, so the system is effectively two-dimensional. The flow
takes place in the surface of a two-dimensional sphere, embedded in
three-dimensional space, and endowed with the metric~\cite{Dub}
\begin{equation}
\label{metric}
ds^{2}=d \rho^{2} + a^{2} \sin^{2} \left (\frac{\rho}{a} \right ) d\varphi^{2},
\end{equation}
where $a$ is the radius of curvature of the sphere, $0 \le \varphi < 2
\pi$ denotes the polar angle measured on the sphere and $0 \le \rho
\le \pi a$ is the geodesic distance from the radial flow source,
located at the sphere's north pole. The sphere has a constant,
positive Gaussian curvature $K=1/a^{2}$ and a constant mean curvature
$H=1/a$. We presume the Gaussian curvature is more relevant than the
mean curvature for reasons discussed in sections~\ref{flat} and
~\ref{nonlinear}. At any stage of our calculation, the ``flat-cell"
limit $a \rightarrow \infty$ (or, equivalently $K \rightarrow 0$)
gives all the well-known familiar results for flow in flat Hele-Shaw
cells.

Denote the viscosities of the upper and lower fluids, 
respectively as $\eta_{1}$ and $\eta_{2}$. Between the two fluids 
there exists a surface tension $\sigma$ (figure 1). 
The flows are assumed to be irrotational, 
except at the interface. Fluid 1 is injected into fluid 2 
through an inlet located at the 
sphere's north pole, at a given flow rate $Q$, which is 
the area covered per unit time. Fluid 2 is simultaneously 
withdrawn, at the same rate, through an outlet placed at the south pole. 

During the flow, the fluid-fluid interface has a perturbed shape
described as $\rho={\cal R} \equiv R + \zeta(\varphi,t)$. The
interface perturbation amplitude is represented by $\zeta(\varphi,t)$,
and $R=R(t)$ denotes the time-dependent unperturbed radius. We can
calculate $R(t)$ from the time-dependent surface area ${\cal A}(t)=4\pi a^{2}
\sin^{2} (R(t)/2a)$. For flow rate $Q$ we write ${\cal A}(t)=
4\pi a^{2} \sin^{2} (R_{0}/2a) + Qt$ then solve for
\begin{equation}
\label{R}
R(t)=a \arccos \left ( C_{0} - \frac{Qt}{2 \pi a^{2}} \right ),
\end{equation}
where $C_{0}=\cos (R_{0}/a)$, and $R_{0}$ is the unperturbed radius at
$t=0$. The unperturbed shape is a polar cap of geodesic radius
$\rho=R$, surface area ${\cal A}$ and circumference ${\cal L}=2\pi a
\sin{(R/a)}$. Note the identity $Q=v {\cal L}$ where $v=dR/dt$ is the velocity
of the unperturbed interface. 

We express the net perturbation $\zeta(\varphi,t)$ in the form of 
a Fourier expansion
\begin{equation}
\label{z}
\zeta(\varphi,t)=\sum_{n=-\infty}^{+\infty} \zeta_{n}(t) \exp{(i n \varphi)}, 
\end{equation}
where
\begin{equation}
\zeta_{n}(t) = \frac{1}{2\pi} \int_{0}^{2\pi} \zeta(\varphi,t) \exp{(-in\varphi)} ~{\rm d}\varphi
\end{equation}
denotes the complex Fourier mode amplitudes and $n$=0, $\pm 1$, $\pm
2$, $...$ is the discrete azimuthal wave number.  To keep the area of
the perturbed shape independent of the perturbation $\zeta$, we
express the Fourier mode corresponding to $n=0$ in the
expansion~(\ref{z}) as
\begin{equation}
\label{z0}
\zeta_{0}(t)= -\frac{1}{2a} \cot{\left ( \frac{R}{a} \right )} \sum_{n \neq 0} |\zeta_{n}(t)|^{2}.
\end{equation}
The constraint~(\ref{z0}) is intrinsically a nonlinear concern 
and is not required in linear stability analysis. 

Since we are interested in both linear and early nonlinear behavior 
of the system, we must derive a differential equation 
for $\zeta_{n}$, correct to second order. This second-order mode-coupling 
equation considers the presence of a full spectrum of modes. We begin the hydrodynamic study of the system by considering a 
generalized version of the usual Darcy's law~\cite{Saf,Rev}, 
adjusted to describe flow between concentric spheres 
(see Appendix~\ref{app})
\begin{equation}
\label{Darcy}
{\bf v}_{j}= - \frac{b^{2}}{12\eta_{j}}~{ \nabla} p_{j},
\end{equation}
where ${\bf v}_{j}={\bf v}_{j}(\rho,\varphi)$ and 
$p_{j}=p_{j}(\rho,\varphi)$ are, 
respectively, the velocity and pressure in fluids 
$j=1$ and $2$. The gradient in equation~(\ref{Darcy}), associated 
with the metric~(\ref{metric}), is~\cite{Dub}
\begin{equation}
\label{gradient}
{ \nabla}=\frac{\partial}{\partial \rho} ~{\bf \hat{\rho} } + \frac{1}{a \sin(\rho/a)} \frac{\partial}{\partial \varphi} 
~ {\bf \hat{\varphi} },
\end{equation}
where the unit vectors $\hat{\bf \rho}$ and $\hat{\bf \varphi}$ 
point in the direction of increase of $\rho$ and $\varphi$, 
respectively. Equation~(\ref{gradient}) was
obtained from the corresponding three-dimensional expression for the 
gradient in spherical coordinates $(r,\theta,\varphi)$, by keeping $r=a$ 
and noting that $\theta=\rho/a$. 

At the interface, the pressure difference between the two fluids is
governed by the mean curvature (the average of the two principal
curvatures) of the fluid-fluid interface~\cite{Rev}.  We can identify
the directions of the principal curvatures in the limit of $b$ smaller
than any other length scale by considering a ``tangent Hele-Shaw
cell'' consisting of two parallel planes tangent, respectively to the
inner and outer spheres at some point along the interface. Within the
tangent Hele-Shaw cell, one principal curvature is associated with
the interface profile in the direction perpendicular to the tangent
planes. We call this curvature $\kappa_{\bot}$ and note that it
is of order $1/b$ with a specific value set by interface contact angles.
The remaining direction of principal curvature is parallel to the tangent plane
and tangent to the interface. We call this curvature $\kappa_{\|}$. Then
the pressure jump boundary condition may be written
\begin{equation}
\label{pressure}
(p_{1} - p_{2})|_{{\cal R}} = 
\sigma \left ( \kappa_{\|} + \kappa_{\bot} \right )|_{{\cal R}}.
\end{equation}
As was the case 
for flow in flat Hele-Shaw cells, $\kappa_{\bot}$ is much
larger than $\kappa_{\|}$ but is is nearly
constant~\cite{McL,Par}. This curvature does not significantly affect
the motion in our problem, because its gradient is nearly zero.

Since the closed boundary describing the fluid-fluid interface is 
itself on the top of a curved surface (sphere), the calculation of 
the ``intra-surface" interface curvature $\kappa_{\|}$ is not as simple as 
it was in the flat-cell case~\cite{Wil,Dub}. Taking into consideration 
the fact that the interface evolves in the surface 
of a sphere of radius $a$, we derive a slightly involved expression 
for the fluid-fluid interface curvature
\begin{equation}
\label{parallel}
\kappa_{\|}=\frac{ \left [ \cos(\rho/a)~a^2\sin^{2}(\rho/a) + 2 \cos(\rho/a) \left ( \frac{\partial \rho}{\partial \varphi} \right )^2 - a \sin(\rho/a) ~\frac{\partial^2 \rho}{\partial \varphi^{2}} \right ]}{ \left [ a^2 \sin^{2}(\rho/a) + \left (\frac{\partial \rho}{\partial \varphi} \right )^2 \right ]^{3/2}}.
\end{equation}
The sign convention for the curvature $\kappa_{\|}$ is such that a
circular interface above the equator has positive curvature, whereas
it has negative curvature below the equator. Keeping terms up to second 
order in the perturbation 
amplitude $\zeta$, we rewrite the interfacial 
curvature as
\begin{equation}
\label{kappa}
(\kappa_{\|})|_{{\cal R}}=\left \{ \frac{C}{aS} - \frac{1}{a^{2}S^{2}} \left ( \zeta + \frac{\partial^2 \zeta}{\partial \varphi^{2}} \right ) + \frac{C}{a^{3}S^{3}} \left [ \zeta^{2} + \frac{1}{2}\left ( \frac{\partial \zeta}{\partial \varphi} \right )^2 + 2 \zeta \frac{\partial^2 \zeta}{\partial \varphi^{2}} \right ] \right \},
\end{equation}
where we have introduced the shorthand notation $S=\sin(R/a)$ and
$C=\cos(R/a)$. Exactly at the equator ($C=0)$, only the term
in~(\ref{kappa}) that is linear in $\zeta$ survives.

Taking advantage of the irrotational and incompressible flow conditions, 
we define the velocity potential $\phi_{j}$ in each of the fluids, 
where ${\bf v}_{j}=-{\bf \nabla} \phi_{j}$. The velocity potential 
satisfies Laplace's equation ${\bf \nabla}^{2}\phi_{j}=0$, where 
${\bf \nabla}^{2}$ is 
the two-dimensional Laplacian defined on the surface of the sphere. 
Combining the velocity potential with equations~(\ref{pressure}) 
and~(\ref{parallel}) for the pressure difference and the generalized 
Darcy's law~(\ref{Darcy}),
we write the equation of motion
\begin{equation}
\label{dimensionless2}
A \left ( \frac{\phi_{1}|_{{\cal R}} + \phi_{2}|_{{\cal R}}}{2} \right ) -  \left ( \frac{\phi_{1}|_{{\cal R}} - \phi_{2}|_{{\cal R}}}{2} \right ) = - \alpha ~(\kappa_{\|})|_{{\cal R}},
\end{equation}
where
\begin{equation}
\label{contrast}
A=\frac{\eta_{2} - \eta_{1}}{\eta_{2} + \eta_{1}}
\end{equation}
is the viscosity contrast and  
\begin{equation} 
\label{alpha}
\alpha=\frac{b^{2} \sigma}{12(\eta_{1} + \eta_{2})},
\end{equation}
contains the surface tension.

Now define Fourier expansions for the velocity potentials $\phi_{j}$.
Far from the interface the velocity field should approach the
unperturbed steady flow with a circular interface of radius R. Thus
for $ \rho \rightarrow 0$ and $\rho \rightarrow \pi a$ the velocity
potentials $\phi_{j}$ approach $\phi_{j}^{0}$, the velocity potentials
for purely radial ($\hat{\bf \rho}$ direction) flow, satisfying
Laplace's equation
\begin{equation}
\label{steady}
\phi_{j}^{0}=- \frac{Q}{2 \pi} 
\log{\left [ \frac{ \tan (\rho/2a)}{\tan(R/2a)}     \right ]} + D_{j},
\end{equation}
where $D_{j}$ are independent of $\rho$ and $\varphi$. The general 
velocity potentials obeying all these requirements are 
\begin{equation}
\label{phi2}
\phi_{j}= \phi_{j}^{0} + \sum_{n \neq 0} \phi_{j n}(t) \left 
[ \frac{ \tan (R/2a)}{\tan(\rho/2a)} \right ]^{(-1)^{j}~|n|} \exp(i n \varphi).
\end{equation}
The trigonometric dependence on $\rho$ transforms, in the flat-cell
limit $a \rightarrow \infty$, into the ratio $R/\rho$.  In order to
calculate the mode coupling differential equation for the system, we
substitute expansions~(\ref{steady}) and ~(\ref{phi2}) into the
equation of motion~(\ref{dimensionless2}), keep second order terms in
the perturbation amplitudes, and Fourier transform them.

To conclude our derivation we need additional relations expressing the
velocity potentials in terms of the perturbation amplitudes.  To find
these, consider the kinematic boundary condition which states that the
normal components of each fluid's velocity at the interface equals the
velocity of the interface itself~\cite{Ros}. Using the
gradient~(\ref{gradient}) we write the kinematic boundary condition
for flow in a sphere as
\begin{equation}
\label{b.c.}
\frac{\partial {\cal R}}{\partial t}= 
\left [ \frac{1}{a^{2} \sin^{2}(\rho/a)} 
\frac{\partial {\rho}}{\partial \varphi}
\frac{\partial \phi_{j}}{\partial \varphi} \right ]_{\rho={\cal R}} 
- \left (\frac{\partial \phi_{j}}{\partial \rho} \right )_{\rho={\cal R}}.
\end{equation}
Inserting expression ${\cal R}= R + \zeta(\varphi,t)$ 
and equation~(\ref{phi2}) for $\phi_{j}$ into the kinematic boundary 
condition~(\ref{b.c.}), we solve for $\phi_{jn}(t)$ consistently 
to second order in $\zeta$ to find 
\begin{eqnarray}
\label{phi1t}
\phi_{1n}(t) & = & -\frac{aS}{|n|}~\dot{\zeta}_{n} - 
\frac{QC}{2 \pi a S |n|} ~\zeta_{n} \nonumber \\
             & + & \sum_{n' \neq 0} \left ( sgn(nn') - 
\frac{C}{|n|} \right ) \dot{\zeta}_{n'}\zeta_{n - n'} + 
\frac{QC}{2 \pi a^{2}S^{2}} \sum_{n' \neq 0} 
\left ( sgn(nn') + \frac{S^{2}}{2C|n|} \right ) 
\zeta_{n'}\zeta_{n - n'}, \nonumber \\
\end{eqnarray}
and
\begin{eqnarray}
\label{phi2t} 
\phi_{2n}(t) & = & \frac{aS}{|n|}~\dot{\zeta}_{n} + 
\frac{QC}{2 \pi a S |n|} ~\zeta_{n} \nonumber \\
             & + & \sum_{n' \neq 0} \left ( sgn(nn') + 
\frac{C}{|n|} \right ) \dot{\zeta}_{n'}\zeta_{n - n'} + 
\frac{QC}{2 \pi a^{2}S^{2}} \sum_{n' \neq 0} 
\left ( sgn(nn') - \frac{S^{2}}{2C|n|} \right ) 
\zeta_{n'}\zeta_{n - n'}. \nonumber \\
\end{eqnarray}
The overdot denotes total time derivative. The sign function $sgn(nn')=1$ 
if $(nn') > 0$ and $sgn(nn')=-1$ if $(nn') < 0$.

We can use relations~(\ref{phi1t}) and~(\ref{phi2t}) to replace the
velocity potentials $\phi_{j}$ in the equation of
motion~(\ref{dimensionless2}) with the perturbation $\zeta$ and its
time derivative $\dot{\zeta}$.  Keeping only quadratic terms in the
perturbation amplitude, and equating Fourier modes $n$ on each side of
equation~(\ref{dimensionless2}), leads to the differential equation
for perturbation amplitudes $\zeta_{n}$. For $n \neq 0$,
\begin{equation}
\label{result}
\dot{\zeta}_{n}=\lambda(n) ~\zeta_{n} + 
\sum_{n' \neq 0} \left [ F(n, n') ~\zeta_{n'} \zeta_{n - n'} + 
G(n, n') ~\dot{\zeta}_{n'} \zeta_{n - n'} \right ],
\end{equation}
where
\begin{equation}
\label{growth}
\lambda(n)= \left [ \frac{Q}{2 \pi a^{2}S^{2}} 
\left (A |n| - C \right ) - \frac{\alpha}{a^{3}S^{3}} |n| (n^{2} - 1) \right ]
\end{equation}
is the linear growth rate, and
\begin{equation}
\label{F}
F(n, n')=\frac{|n|}{aS} \left \{ \frac{QAC}{2 \pi a^{2}S^{2}} 
\left [ \frac{1}{2} - sgn(nn') \right ] - \frac{\alpha C}{a^{3}S^{3}} 
\left [ 1 - \frac{n'}{2} ( 3 n' + n )   \right ] \right \} + 
\frac{Q}{4 \pi a^{3} S},
\end{equation}
\begin{equation}
\label{G}
G(n, n')=\frac{1}{aS} \left \{ A|n| [ 1 - sgn(nn') ] - C \right \}
\end{equation}
are the second-order mode coupling terms. Equation~(\ref{result}) is 
the mode coupling equation of the Saffman-Taylor problem in 
a spherical Hele-Shaw cell. It gives us the time evolution of the 
perturbation amplitudes $\zeta_{n}$ accurate to second order. 
In the following sections we study equation~(\ref{result}) in more 
detail, and investigate the role played by cell geometry in 
the interface dynamics.

\section{Discussion}
\label{discussion}

We use the mode coupling equation~(\ref{result}) to investigate the
linear instability of individual modes and the coupling of a small
number of modes. The most noteworthy effect of curvature we identify
concerns its influence on finger tip-splitting.  Tip-splitting is
related to the influence of a fundamental mode on the growth of its
harmonic~\cite{MW}. We take $n$ as the fundamental and $2n$ as the
harmonic. To observe interfacial instability of the fundamental mode
$n$, we must have $\lambda(n)>0$. This occurs if the destabilizing
contribution $QA$ in Eq.~(\ref{growth}) is positive and sufficiently
large compared with the stabilizing surface tension term proportional
to $\alpha$. To observe growth of the harmonic mode $2n$, we presume
that $QA$ is sufficiently large that $\lambda(2n)$ is non-negative.

To simplify our discussion it is convenient to rewrite the net
perturbation~(\ref{z}) in terms of cosine and sine modes
\begin{equation}
\label{sincos}
\zeta(\theta,t)= \zeta_{0} + 
\sum_{n = 1}^{\infty} \left[ a_{n}(t)\cos(n\theta) + 
b_{n}(t)\sin(n\theta) \right ],
\end{equation}
where $a_{n}=\zeta_{n} + \zeta_{-n}$ and $b_{n}=i \left ( \zeta_{n} -
\zeta_{-n} \right )$ are real-valued. Without loss of generality we
may choose the phase of the fundamental mode so that $a_{n} > 0$ and
$b_{n}=0$.  We replace the time derivative terms $\dot{a}_{n}$ and
$\dot{b}_{n}$ by $\lambda(n)~a_{n}$ and $\lambda(n)~b_{n}$,
respectively, for consistent second order expressions. Under these
circumstances the equations of motion become
\begin{equation}
\label{new3}
\dot{a}_{2n}=\lambda(2n)~a_{2n} + \frac{1}{2} ~T(2n,n) ~a_{n}^2
\end{equation}
\begin{equation}
\label{new4}
\dot{b}_{2n}=\lambda(2n)~b_{2n},
\end{equation}
where the tip-splitting function is defined as
\begin{equation}
\label{T(2n,n)}
T(2n,n)=\left [ F(2n, n) + \lambda(n) ~G(2n, n) \right ].
\end{equation}

Note that the sign of $T(2n,n)$ dictates if finger tip-splitting is
favored or not by the dynamics. If $T(2n,n)<0$, at second order the
result is a driving term of order $a_{n}^{2}$ forcing growth of
$a_{2n} < 0$.  With this particular phase of the harmonic forced by
the dynamics, the $n$ outwards-pointing fingers of the fundamental
mode $n$ tend to split. In this case the driving term in equation of
motion~(\ref{new3}) spontaneously generates the harmonic mode. In
contrast, if $T(2n,n)>0$ growth of $a_{2n} > 0$ would be favored,
leading to outwards-pointing finger tip-sharpening.  Note that mode
$b_{2n}$, whose growth is uninfluenced by $a_{n}$, skews the fingers
of mode $n$.  In the presence of $a_{2n} < 0$, the role of $b_{2n}$ is
to favor one of the two split fingers over the other.

\subsection{The flat-cell limit}
\label{flat}

We begin our discussion by analyzing the flat-cell limit of the
mode-coupling expression~(\ref{result}). We hold fixed the unperturbed
interface velocity $v=Q/{\cal L}$.  Three distinct flat space
limits can be taken: (i) ({\em north pole}) let $a \rightarrow \infty$
holding $R$ finite so that $C \rightarrow 1$ and $aS \rightarrow R$;
(ii) ({\em south pole}) let $a \rightarrow \infty$ holding $R' \equiv
\pi a - R$ finite; so that $C \rightarrow -1$ and $aS \rightarrow R'$;
(iii) ({\em tropical}) let $R/a$ be constant and hold fixed $k \equiv
2\pi n/{\cal L}$.  These three limits correspond to three physically
distinct flat space flow problems. Our goal in this section is to
verify that the linear and nonlinear terms in the equation of
motion~(\ref{result}) reduce to their expected forms in the flat space
limit.  The actual evolution of interfaces according to the equations
of motion is then discussed in greater detail in sections~\ref{linear}
and~\ref{nonlinear}.

At the north pole (case (i), $C \rightarrow 1$), we recover the mode-coupling
equations~\cite{MW} of flat, radial {\it divergent} flow~\cite{Thome},
related to outward radial motion in which fluid 1 pushes fluid
2:
\begin{eqnarray}
\lambda(n)= \left [ \frac{Q}{2 \pi R^{2}} \left (A |n| - 1 \right ) - 
\frac{\alpha}{R^{3}} |n| (n^{2} - 1) \right ] \\ \nonumber
F(n, n')=\frac{|n|}{R} \left \{ \frac{QA}{2 \pi R^{2}} 
\left [ \frac{1}{2} - sgn(nn') \right ] -
\frac{\alpha}{R^{3}} 
\left [ 1 - \frac{n'}{2} ( 3 n' + n )  \right ] \right \} \\
G(n, n')=\frac{1}{R} \left \{ A|n| [ 1 - sgn(nn') ] - 1 \right \}. \nonumber
\end{eqnarray}
Provided the viscosity contrast $A > 0$, so that the less viscous
fluid pushes the more viscous fluid, the interface is linearly unstable
and exhibits finger growth. As the interface perturbation grows, the
nonlinear mode coupling broadens and splits the outward-pointing
fingers and sharpens the inward-pointing fingers. This occurs because
$T(2n,n)<0$ when $\lambda(2n) \ge 0$.

In contrast, at the south pole (case (ii), $C \rightarrow -1$), we obtain flat,
radial {\it convergent} flow~\cite{Thome} equivalent to the inward
radial motion corresponding to withdrawal of fluid 2 surrounded by
fluid 1. Provided that $QA>0$, so that the less viscous fluid displaces
the more viscous fluid, the interface remains linearly
unstable. However, now the outwards-pointing fingers sharpen because
for negative $C$ we find $T(2n,n)>0$ when $\lambda(2n)\ge 0$. The
asymmetry between north and south pole behaviors occurs primarily in
the nonlinear term and comes from the terms proportional to $C$.  In
contrast, the linear growth rate $\lambda(n)$ is nearly symmetric between
the north and south pole limits. This can be understood because
interchanging the north and south poles is equivalent to reversing the
sign of $Q$ (the direction of flow) and the sign of $A$ (interchanging
the fluid viscosities) while holding the surface tension $\alpha$
unchanged. However, the term in $\lambda(n)$ proportional to $C$ breaks
this symmetry slightly. 

A special example of the tropical case (iii) is the equatorial limit,
for which $C=0$ and $S=1$. This case should be compared with the
cylindrical flow geometry, taking the cylinder tangent to the sphere
at the equator. The flat space limit of this problem indeed reduces to
the problem of rectangular flow in flat space with periodic boundary
conditions. In particular,
\begin{eqnarray}
\lambda(n) \rightarrow |k| \left[ A v - \alpha k^2 \right] \\ \nonumber
F(n,n') \rightarrow 0 \\
G(n,n') \rightarrow A |k| \left[ 1 - sgn (kk') \right]. \nonumber
\end{eqnarray}
As expected, there is no tendency for tip-splitting on the cylinder.
This alludes to our suggestion that tip-splitting is controlled by
Gaussian curvature $K$ rather than mean curvature $H$ because
variation of the radius of curvature of the cylinder alters the mean
curvature while the Gaussian curvature remains zero. Tip-splitting is
absent for any value of the mean curvature.

Before the limit is reached, there are small differences between the
spherical and cylindrical cases. For example, the surface tension
contribution to $\lambda(n)$ vanishes for $k=0$ on a cylinder
corresponding to translation invariance of the interface length. On a
sphere the corresponding displacement moves the interface from the
equator to the tropics, shortening the interface and lowering the
surface energy.  However, on a sphere the modes $n= \pm 1$ correspond
to a global off-center shift of the circular interface preserving
circular shape and perimeter. Thus the surface tension term vanishes
for $n=\pm 1$ on the sphere while this mode increases the
perimeter and raises the energy on a cylinder.

\subsection{Linear growth}
\label{linear}

Consider the purely linear contribution, which appears as the first
term on the right hand side of equation~(\ref{result}).  Since $R$
varies with time, the linear growth rate $\lambda(n)$ is time
dependent as well. This implies that the actual relaxation or growth
of mode $n$ is not proportional to the factor $\exp[\lambda(n)t]$, but
rather
\begin{equation}
\label{relax}
\zeta_{n}(t)=\zeta_{n}(0) \exp \left [ \int_{0}^{t} \lambda(n)~{\rm d}t' \right].
\end{equation}
If $\int_{0}^{t} \lambda(n) ~{\rm d}t' > 0$ the disturbance grows, indicating
instability.  Two relevant facts can be extracted from the linear
growth rate: (i) the existence of a series of critical radii
$R_c(n)$ (defined by setting $\lambda(n)=0$) at which the interface
becomes unstable for a given mode $n$; (ii) the presence of a fastest
growing mode $n^{\ast}$, given by the closest integer to the maximum
of equation~(\ref{growth}) with respect to $n$ (defined by setting
$d\lambda(n)/dn=0$).  In view of equation~(\ref{relax}) $n^{\ast}$ is
not simply related to the number of fingers present, even in the early
stages of pattern formation.  Furthermore, in the nonlinear regime the
subsequent tip-splitting process and mode competition result in a
final number of fingers which can differ from the number present in
the linear regime.

Cardoso and Woods~\cite{Car} analyze flat space radial flow assuming
the presence of a constant low level of noise during the whole
evolution of the interface (their ``model B'', also see
Ref.~\cite{MW}). The sources of noise may come from
inhomogeneities on the surface of the Hele-Shaw cell, irregularities
in the gap thickness $b$, or from thermal or pressure
fluctuations~\cite{Gin}. The predictions of this model are in
qualitative agreement with experimental observations within the linear
regime~\cite{Car} and the nonlinear regime~\cite{MW}.

Suppose that we begin with an initially circular interface that is
steadily expanding.  During the interface expansion each mode $n$ is
perturbed with a constant (in time) random complex amplitude
$\zeta_{n}(0)$.  This noise amplitude contains an $n$ dependent random
phase but its magnitude $|\zeta_{n}(0)|$ is independent of $n$ by
assumption. As the interface continues to expand, it progressively
reaches critical radii $R_{c}(n)$ for $n=$ $2$ , $3$ , $...$ .  Once a
particular $R_{c}(n)$ is reached, the perturbation amplitude
$\zeta_{n}$ starts to vary with time.  Within this model, the first
order (linear) solution of equation~(\ref{result}) can be written as
\begin{equation}
\label{linearsol}
\zeta_{n}^{lin}(t)=\left\{ \begin{array}{cl}
\zeta_{n}(0) &\mbox{if $R < R_{c}(n)$} \\
\zeta_{n}(0) \left \{ \left ( \frac{1 + C_{c}}{1 + C} \right ) 
\left [ \frac{1}{{\cal T}_{2}} \right ]^{A|n| - 1} \exp \left [ C(A|n| - C) 
\left ( {\cal T}_{1} - 1 \right ) \right ] \right \} &\mbox{if $R \geq R_{c}(n)$}
\end{array}\right.
\end{equation}
where the functions ${\cal T}_{1}=\tan{[R_{c}(n)/a]}/\tan{(R/a)}$, 
${\cal T}_{2}=\tan{[R_{c}(n)/2a]}/\tan{(R/2a)}$, 
and $C_{c}=\cos{[R_{c}(n)/a]}$.

To see the overall effect of Eq.~(\ref{linearsol}), we plot the time
evolution of the interface using the experimental parameters given in
Paterson's classical experiment~\cite{Pat}.  Paterson observed the
rapid growth of fingers, as air ($\eta_{1} \approx 0$) was blown at a
relatively high injection rate, $Q=9.3$ $\rm{cm^{2}/s}$, into
glycerine ($\eta_{2} \approx 5.21$ $\rm{g}$/($\rm{cm}$ $\rm{s}$)) in a
radial source flow Hele-Shaw cell. The thickness of the cell $b=0.15$
$\rm{cm}$ and the surface tension $\sigma=63$ $\rm{dyne/cm}$.  We take
into account modes $n$ ranging from $n=2$ up to $20$.  We evolve from
initial radius $R_{0}=0.05$ $\rm{cm}$. The noise amplitude
$|\zeta_{n}(0)|=R_{0}/1000$.  Figure 2 depicts the evolution of the
interface, for a random choice of phases, up to time $t=20$
$\rm{s}$. We set the radius of the sphere $a=10$ $\rm{cm}$. We encourage the
reader to compare the resulting interface with equivalent figures in
flat space. Ref.~\cite{MW} contains a figure in which growth
conditions are similar.  In particular, the {\em phases} of the
initial perturbations are identical, although in the present case we
choose smaller initial amplitudes. It is also of interest to compare
with flat space experimental patterns found in the
literature~\cite{{Bat},{Pat},{Thome},{Car},{Che}}.

\subsection{Nonlinear behavior and tip-splitting}
\label{nonlinear}

To visualize the consequences of the second order term in the
equations of motion, we solve equation~(\ref{result}) to second order
accuracy.  If we substitute the linear solution given in
equation~(\ref{linearsol}) into the second-order terms on the right hand
side of equation~(\ref{result}), we obtain the differential equation
\begin{equation}
\label{try2}
\dot{\zeta}_{n}=\lambda(n)~\zeta_{n} + W(n,t),
\end{equation}
where 
\begin{equation}
\label{W}
W(n,t)=\sum_{n' \neq 0} \left [ F(n, n') ~\zeta_{n'}^{lin} \zeta_{n - n'}^{lin} + G(n, n') ~\dot{\zeta}_{n'}^{lin} \zeta_{n - n'}^{lin} \right ]
\end{equation}
acts as a driving force in the linearized equation of motion~(\ref{try2}).
Equation~(\ref{try2}) is a standard first order linear differential 
equation~\cite{Gra} with the solution
\begin{equation}
\label{sol}
\zeta_{n}(t)=\left\{ \begin{array}{cl}
\zeta_{n}(0) &\mbox{if $R < R_{c}(n)$}\\
\zeta_{n}^{lin}(t) \left \{ 1 + \int_{t_{c}(n)}^{t} \left [ \frac{W(n,t')}{\zeta_{n}^{lin}(t')} \right ] dt' \right\} &\mbox{if $R \geq R_{c}(n)$}.
\end{array}\right.
\end{equation}
Here $t_{c}(n)$ is the time required for the unperturbed growth 
to reach radius $R_{c}(n)$ and can be calculated from equation~(\ref{R}).
This solution describes the weakly nonlinear evolution, where the dominant 
modes just become coupled by nonlinear effects. 

We use the second order solution~(\ref{sol}) to investigate the
nonlinear coupling among various modes $n$.  In figure 3, we plot the
interface for a certain time ($t=20$ ${\rm s}$), considering the same
random choice of initial phases as was employed in figure 2, and
coupling all modes with $2 \le n \le 20$.  The nonlinear evolution
leads to wider fingers and their tips become more blunt. These fingers
spread and some of them start to bifurcate, by splitting at the
tip. As was shown in Eq.~(\ref{new3}), tip splitting is caused by
fingers of mode $n$ driving the growth of their own harmonic $2n$.

Now we turn our attention to the investigation of how curvature
influences finger tip-splitting for Hele-Shaw flow on a sphere.  It
has been shown in reference~\cite{MW} that, for flat, divergent radial
flow under the condition that $\lambda(2n) \ge 0$, we have
$T(2n,n)<0$.  Thus, when the harmonic is able to grow, finger
tip-splitting is favored. In section~\ref{flat} we showed that in the
limit of small cell curvature $K$ we recover the flat space limit
where we know $T(2n,n)$ is negative. To carry out our current
analysis, we consider the case in which $\lambda(2n)=0$ and see how
the curvature influences $T(2n,n)$.

Since spherical Hele-Shaw flow involves many independent parameters we
change only one relevant quantity at a time to see what each one
does. We hold the unperturbed interface velocity $v$ and the
unperturbed interface contour length ${\cal L}$ fixed, while varying the
curvature $K$. This isolates the influence of spatial curvature $K$
from the effect of variations in $v$ and ${\cal L}$.  Fixed ${\cal L}$
means that fixed mode number $n$ corresponds to a fixed wavelength.

We adopt an {\it instantaneous} approach: we look at the linear growth
rate and mode coupling at an instant in time, ignoring the past
history of how a given interface arose from some initial condition
followed by growth. Such an instantaneous approach, at which ${\cal L}$
and $v$ have a particular value, enables us to compare the behavior of
interfaces evolving in distinct background curvatures, but under
dynamically equivalent circumstances. Moreover, at the instant when
the interface circumference ${\cal L}$ in curved space equals the
circumference in flat space, if the two velocities also match, the
identity $Q=v {\cal L}$ shows that the $Q$ value in flat space equals
the $Q$ value in curved space.  Therefore, it is advantageous to look
at the instantaneous tendency towards tip-splitting for interfaces of
identical unperturbed $v$ and ${\cal L}$.

Consider a particular $v$ and ${\cal L}$ combination at the onset of
growth of mode $2n$ (using the condition $\lambda(2n)=0$) in the limit
of flat space, where it is known that $T(2n,n)<0$.  To illustrate how
tip-splitting varies with curvature for flow on a sphere, we plot in
figure 4 the tip-splitting function $T(2n,n)$ as a function of $K$.
The solid (dashed) curves refer to the behavior in the northern (southern)
hemisphere. By inspecting figure 4 we notice that, in the northern
hemisphere there is a suppression of tip-splitting for increasingly
larger curvature because the magnitude of $T(2n,n)$ is maximum for the
flat, divergent radial case ($K=0$) and decreases as curvature is
increased.  For each $n$ there is some value of $K$ for which the
circle of circumference ${\cal L}$ hits the equator. This is precisely
the point at which northern and southern hemisphere branches of the
curves meet in figure 4.

For the southern hemisphere $T(2n,n)$ is always positive.
Consequently, we should not expect finger tip-splitting of
outward-pointing fingers in this hemisphere. Actually, finger
tip-splitting is replaced by tip-narrowing, along with a splitting of
the inward-pointing fingers. Finger tip-narrowing is regulated by the
curvature $K$.

Another noteworthy point about figure 4 is the evident
symmetry-breaking in $T(2n,n)$ between northern and southern
hemispheres.  The justification for this asymmetry is similar to the
one that explains why the interface is always in-out asymmetric in
flat radial flow: if we are located at either pole we can always
distinguish regions that are inside and outside the interface. That is
why convergent and divergent flat, radial flows are not
equivalent~\cite{Thome}.  From this point of view, the asymmetry
observed in figure 4 should be somehow expected.

We conclude this section presenting the lowest order curvature expansion 
for $T(2n,n)$ and studying how curvature $K$ influences tip-splitting for 
small curvature values. Using the instantaneous approach described 
above, series expansion of equation~(\ref{T(2n,n)}) for the 
northern hemisphere yields
\begin{equation}
\label{expandedT}
T(2n,n)=\frac{Q^{2}~(2n^{2} + 1)}{32 \pi^{2} \alpha ~n(2n + 1)^2} 
\left [ - \frac{Q^{2}}{8 \pi^{2} \alpha ~n^{2} (2n + 1)^2} + K \right ] 
+ {\cal O} ~(K^{2}).
\end{equation}
The perturbative expansion~(\ref{expandedT}) for small $K$ explicitly
shows a linear correction to the flat space limit ($K=0$) that reduces
the magnitude of $T(2n,n)$, inhibiting finger tip-splitting. A similar
kind of expansion can be done for the southern hemisphere, resulting
in enhanced tip-sharpening. Those results are in agreement with the
small curvature behavior depicted in figure 4.

Equation~(\ref{expandedT}) allows us to make predictions about 
tip-splitting behavior for Hele-Shaw flow in (weakly) negatively curved 
backgrounds. This type of flow could happen, for instance, 
between two saddle-like surfaces. Actually, flow in porous media 
seems to be somehow linked to flow on negatively curved surfaces. Porous 
materials define multiply-connected surfaces, presenting 
negative {\it average} Gausssian curvature
\begin{equation}
\label{mean}
\bar{K}=\frac{\int K d {\cal S}}{\int d {\cal S}}
=\frac{4 \pi ~(1 - g)}{\int d {\cal S}},
\end{equation}
where $d {\cal S}$ denotes an infinitesimal area element and $g$ is 
the so-called genus~\cite{Dub}, which denotes the number of 
holes present in a given surface. Expression~(\ref{mean}) relates 
the integral of the Gaussian curvature $K$ of a given surface to 
its topological properties, by virtue of the Gauss-Bonnet 
theorem~\cite{Dub}. From~(\ref{mean}) 
we verify that $\bar{K}$ becomes progressively negative 
when the number of holes is increased: that is why 
$\bar{K}>0$ for a sphere ($g=0$), $\bar{K}=0$ for 
a torus ($g=1$) and $\bar{K}<0$ for a 
$g$-torus ($g \ge 2$). In this sense, a medium 
which is rich in pores (holes) would present negative curvature 
features. However, flow on surfaces of {\it constant} negative 
Gaussian curvature is complex and to treat the problem rigorously 
would require an interesting generalization of 
our formalism beyond the scope of 
the present paper.

Here we simply point out what would be the behavior 
for flow on surfaces that are slightly (negatively) curved. 
It is easy to see, by performing the substitution 
$K  \rightarrow -K$ in~(\ref{expandedT}), that negative curvature 
should enhance finger splitting in comparison to the flat case. 
This is in striking contrast to the problem of flow in positively 
curved surfaces, such as the sphere, where curvature leads to suppression 
of tip-splitting. This sets an important distinction between Hele-Shaw 
flows on surfaces with negative curvature and those on surfaces with 
positive $K$. Our results confirm explicitly the predictions made 
in reference~\cite{MW} about the role of cell curvature on finger 
splitting.

The fact that the correction is linear in $K$ supports our suggestion
that Gaussian curvature is more relevant than mean curvature. The same
expansion Eq.~(\ref{expandedT}) could be written in terms of the
square of mean curvature $H$.  However, then we would predict
supression of tip-splitting for both positive $(H>0)$ and negatively 
$(H<0)$ curved surfaces. We believe tip-splitting is 
enhanced for negative Gaussian
curvature because the metric creates ample space for the split fingers
to penetrate without mutual competition.

\section{Concluding remarks}
\label{conclude}

In this paper we generalized the traditional Saffman-Taylor problem 
by studying viscous flow on curved surfaces. Our main purpose was 
to investigate the influence of spatial curvature on viscous 
fingering pattern formation, when fluid flow takes place on 
a sphere. By deriving the equation of motion for the interface 
pertubation amplitudes, using a mode-coupling approach, a study 
of both linear and weakly nonlinear stages of evolution could be 
carried out.

We have shown that cell curvature can be used as 
a control parameter to discipline splitting of the viscous 
fingers. The fluid-fluid interface can be more stable 
or unstable, with respect to tip-splitting, depending on the 
curvature of the surface to which the flow is confined. We also 
detected an asymmetry on tip-splitting behavior depending where 
the interface evolves: while tip-splitting may be still present 
on the northern hemisphere, it is completely replaced by finger 
tip-sharpening on the southern hemisphere. We also found 
evidence that Hele-Shaw flows on negatively curved surfaces 
would present enhanced tendency to tip-splitting, so highly 
branched patterns may be expected. In summary, we have explicitly 
verified that interfacial behavior is coupled to the geometry 
of the Hele-Shaw cell, so that curvature has important 
consequences for flow dynamics.

\vspace{0.5 cm}
\begin{center}
{\bf ACKNOWLEDGMENTS}
\end {center}
\noindent
F.P., F.M. and J.A.M. thank CNPq and FINEP (through 
its PRONEX Program) for financial support. Work of M.W. was supported 
in part by the National Science Foundation grant No. DMR-9732567.

\appendix
\section{Darcy's Law on a Sphere}
\label{app}
This appendix derives Darcy's law for viscous flow between concentric
spheres. The derivation is based upon results presented in Bird, {\it
et al.}~\cite{Bird}. We begin with a coordinate-free representation of
the continuity equation for an incompressible fluid
\begin{equation}
\label{cont}
{\nabla} \cdot {\bf u} = 0
\end{equation}
and the Navier-Stokes equation
\begin{equation}
\label{NS-CF}
\rho \left [ {{\partial {\bf u}}\over{\partial t}} 
+ ({\bf u} {\cdot \nabla}) {\bf u} \right ]
= - { \nabla} p + \eta { \nabla}^{2} {\bf u} 
\end{equation}
where ${\bf u}$ denotes the three-dimensional fluid velocity and we
neglect the acceleration due to gravity. Neglecting the inertial terms
on the left-hand side of Eq.~(\ref{NS-CF}), transforming to spherical
coordinates $(r,\theta,\varphi)$, and specializing to the case of {\it
polar} flow $(u_r=u_{\varphi}=0)$~\cite{Bird,Vat}, we rewrite the
Navier-Stokes equation
\begin{equation}
\label{NS-SC}
{{1}\over{r}}{{\partial p}\over{\partial \theta}}
= \eta \left (\nabla^2 u_{\theta} 
- {{u_{\theta}}\over{r^2 \sin^2{\theta}}} \right ).
\end{equation}
The noteworthy term in Eq.~(\ref{NS-SC}) is the second term multiplying
viscosity which enters as a result of the curvilinear coordinate
system.

To determine $u_{\theta}(r,\theta)$ note that the continuity
equation~(\ref{cont}) adapted for polar flow
\begin{equation}
\label{CE}
{{1}\over{r \sin{\theta}}}
{{\partial (u_{\theta} \sin{\theta})}\over{\partial \theta}}
= 0
\end{equation}
demands a solution of the form
\begin{equation}
\label{general}
u_{\theta}(r,\theta)={{u(r)}\over{\sin{\theta}}}.
\end{equation}
Insert this form into the Navier-Stokes equation~(\ref{NS-SC}) 
and multiply by $r \sin{\theta}$ to separate variables:
\begin{equation}
\label{NS-theta}
\sin{\theta}{{\partial p}\over{\partial \theta}}
= {{\eta}\over{r}}
{{\partial}\over{\partial r}}
\left ( r^2 {{\partial u}\over{\partial r}} \right ).
\end{equation}
Because the left-hand side involves only angle $\theta$ and the
right-hand side is radial, involving only $r$, each side must be
constant sharing a common value $B$.  The solution of the radial
equation subject to no-slip boundary conditions $u(a)=u(a+b)=0$ is
\begin{equation}
\label{uofr}
u(r)={{a u_0}\over{b^2}} {{(r-a)(a+b-r)}\over{r}}
\end{equation}
where $B=-2 \eta a u_0/b^2$.

Darcy's law for Hele-Shaw cells is obtained by averaging the
three-dimensional velocity ${\bf u}$ over the gap width $b$. We choose
to average in the {\em radial} direction $r$ and obtain
$\bar{v}_{\theta}=\bar{u}/\sin{\theta}$ with
\begin{equation}
\label{ubar}
\bar{u}={{u_0}\over{6}} {\cal F} \left ( {{b}\over{a}} \right )
\approx {{u_0}\over{6}} \left ( 1-{{b}\over{2a}} + \cdot\cdot\cdot \right),
\end{equation}
where 
\begin{equation}
\label{eff}
{\cal F} \left ( {{b}\over{a}} \right )=6 \left ( \frac{a}{b} \right )^{3} \left [ \frac{b}{a} \left ( 1 + \frac{b}{2a} \right ) - \left ( 1 + \frac{b}{a} \right ) \log{\left ( 1 + \frac{b}{a} \right )} \right ].
\end{equation}
Equation~(\ref{eff}) could also be written, in a more compact form, as a 
hypergeometric function ${\cal F}(b/a)=F(1,2;4;-b/a)$.

Equation~(\ref{ubar}) generalizes the usual flat-cell 
velocity average. Darcy's law becomes
\begin{equation}
\label{newDarcy}
\bar{v}_{\theta} = 
- {{b^2 {\cal F} \left( {{b}\over{a}} \right )}\over{12\eta}}
\left ( {{1}\over{a}} {{\partial p}\over{\partial \theta}} \right).
\end{equation}
The effect of curvature can thus be incorporated entirely into a
reduced gap width $b$ or an enhanced viscosity $\eta$.
Equation~(\ref{newDarcy}) recovers the usual Darcy's law for flow in
flat Hele-Shaw cells with corrections of higher order in $b/a$.  In
this work we will be interested in the case $b \ll \zeta \ll \rho$,
with $\rho \sim a$.

\pagebreak
\noindent
\centerline{{\large {FIGURE CAPTIONS}}}
\vskip 0.5 in
\noindent
{FIG. 1:} Schematic configuration of the flow in a spherical Hele-Shaw 
cell.
\vskip 0.5 in
\noindent
{FIG. 2:} Time evolution of the fluids according to Eq.~(\ref{linearsol}),
including modes $2 \le n \le 20$. The initial perturbation amplitudes
$| \zeta_n(0) |=R_0/1000$ and $R_0=0.05$ $\rm{cm}$. Other parameters are given 
in the text. We show the fluid-fluid interface at $t=20$ $\rm{s}$.

\vskip 0.5 in
\noindent
{FIG. 3:} Nonlinear evolution according to Eq.~(\ref{sol}). All physical
parameters and initial conditions are the same as those used in
Fig. 2.

\vskip 0.5 in
\noindent
{FIG. 4:} Variation of $T(2n,n)$ as a function of the 
spherical Hele-Shaw cell curvature $K$, 
for modes $n=8,10,12$. The solid (dashed) curves describe behavior in 
the Northern (Southern) hemisphere. The units of $T(2n,n)$ and $K$ 
are $({\rm cm~s})^{-1}$ and ${\rm cm}^{-2}$, respectively.

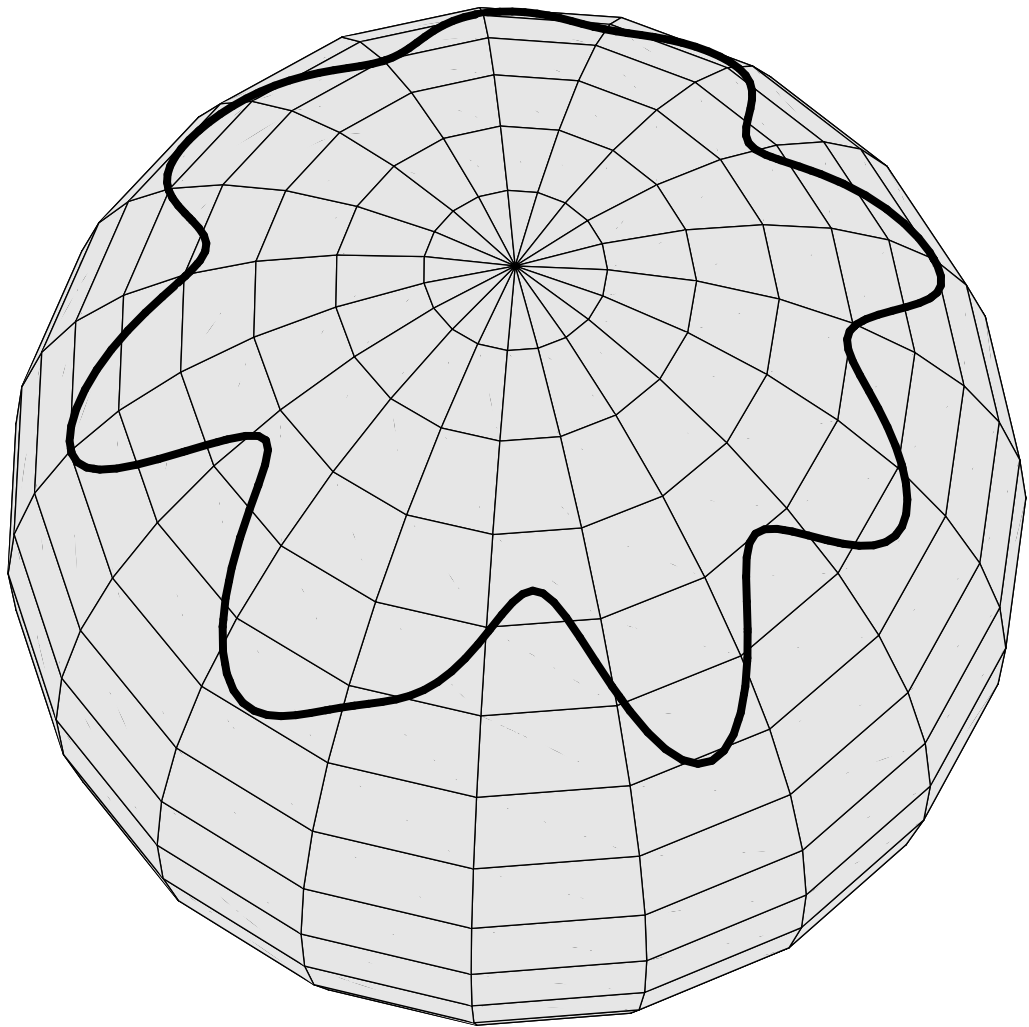
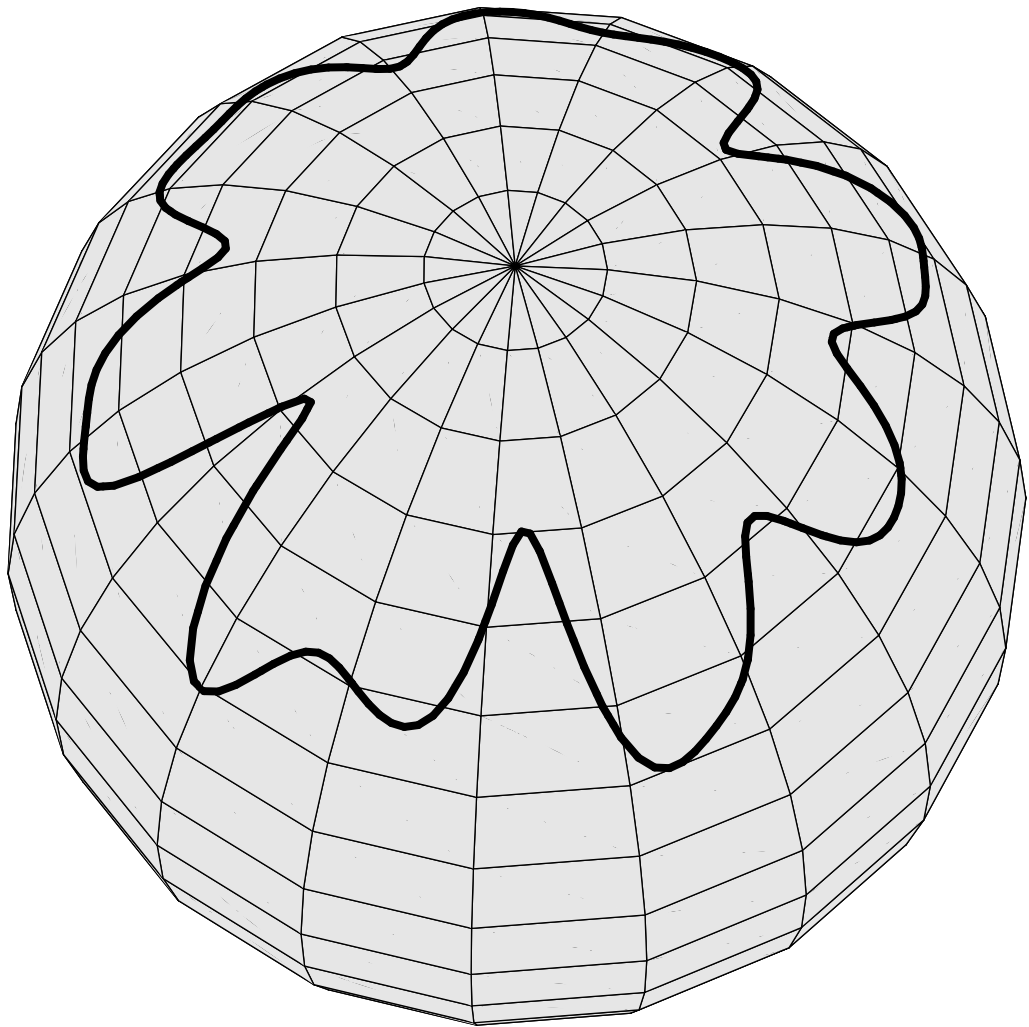
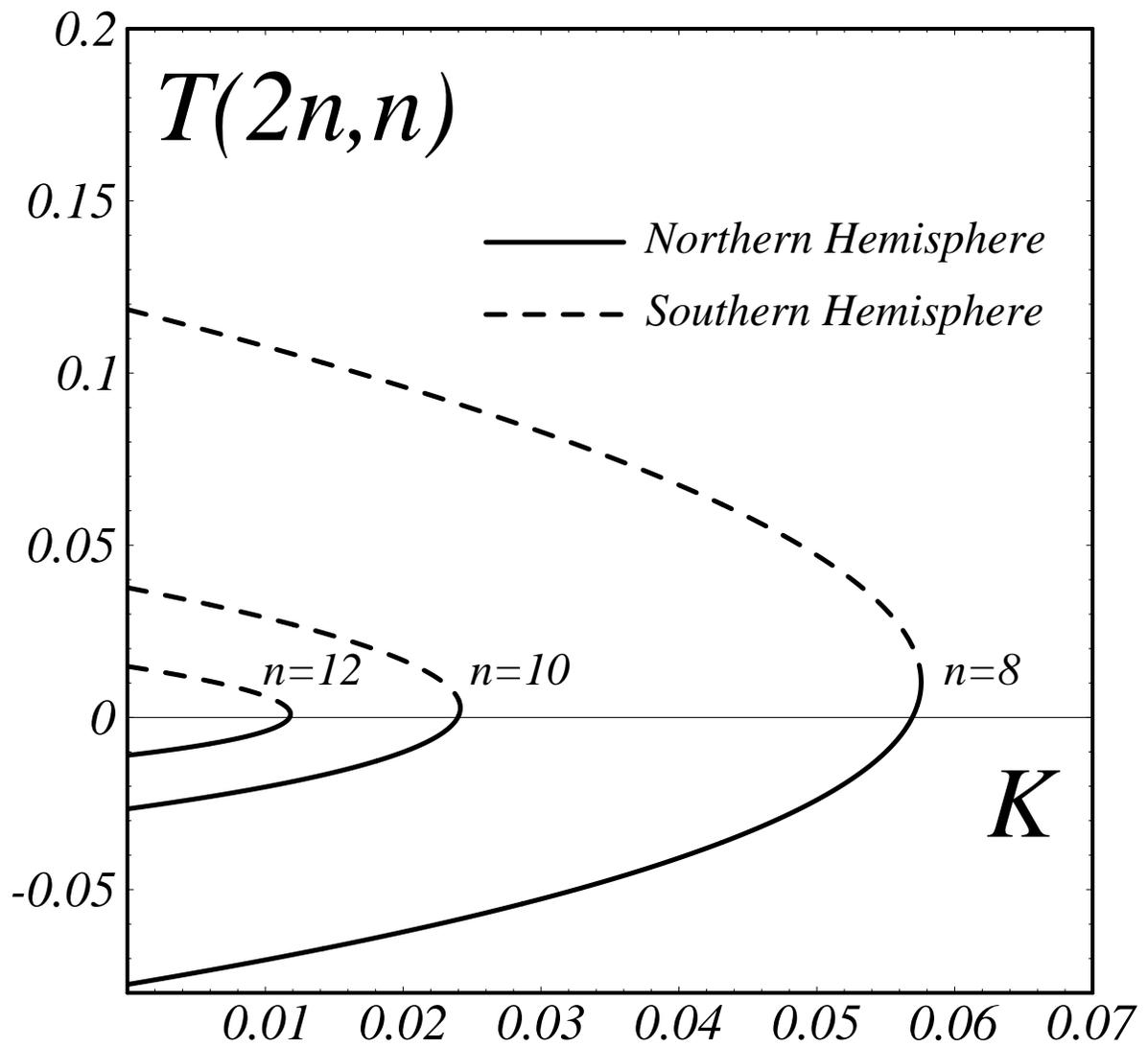
\end{document}